\documentclass[twocolumn,showpacs,preprintnumbers,amsmath,amssymb]{revtex4}
\usepackage[dvips]{graphicx} % para incluir gráficos
\begin{document}
\preprint{IMAFF-RCA-03-09}
\title{Coherent states in the quantum multiverse}

\author{S. Robles-P\'{e}rez$^{1,2}$, Y. Hassouni$^{3}$ and P. F.
Gonz\'{a}lez-D\'{i}az$^{1,2}$} \affiliation{$^{1}$Colina de los
Chopos, Centro de F\'{\i}sica ``Miguel Catal\'{a}n'', Instituto de
F\'{\i}sica Fundamental,\\ Consejo Superior de Investigaciones
Cient\'{\i}ficas, Serrano 121, 28006 Madrid (SPAIN).
\\$^2$Estaci\'{o}n Ecol\'{o}gica de Biocosmolog\'{\i}a,
Medell\'{\i}n (SPAIN).
\\$^{3}$Laboratoire de Physique Theorique, Facult\'{e} des
Sciences-Universit\'{e} Sidi Med Ben Abdellah, \\ Avenue Ibn
Batouta B.P: 1014, Agdal Rabat (MOROCCO). }
\date{\today}
\begin{abstract}
In this paper, we study the role of coherent states in the realm
of quantum cosmology, both in a second-quantized single universe
and in a third-quantized quantum multiverse. In particular, most
emphasis will be paid to the quantum description of multiverses
made of accelerated universes. We have shown that the quantum  states
involved at a quantum mechanical multiverse whose single universes
are accelerated are given by squeezed states having no classical analogs.
\end{abstract}


\maketitle

\section{Introduction}

Coherent states have been always considered as rather mathematical
objects with application in quantum physics, and they can also
represent a solid basis for the quantum description of a
particular system \cite{Glauber63}. Therefore, obtaining coherent
states in quantum cosmology will allow us both: i) to enhance the
analogy between usual quantum mechanics and cosmology, and ii) to
prepare the mechanics to describe the universe further, potentially
generalizable developments.

On the other hand, coherent states can be constructed from the
algebras lying behind their definition. More precisely, in the
literature Heisenberg algebras are usually used to obtain
them. Nevertheless, in some works \cite{Klauder63} coherent states
defined for given quantum systems are constructed from the so-called
Generalized Heisenberg Algebras (GHA). These allow us to construct
coherent states without specifying any formal expressions for the
annihilation operator. Such algebras will be specially useful to describe the case
of a universe in second quantization.

Furthermore, second-quantization of the universe can provide us
with the quantum state of a single universe by means of a
wavefunction \cite{Hartle83}, when given by a pure state, or
through a density matrix \cite{Page86} if, instead, it is more
generally given in terms of a mixed state. However, in any of the
above representations one cannot account for any topology changes \cite{Hawking87},
i. e. the creation or annihilation of universes. Therefore, a
third-quantization procedure is needed to quantum mechanically
describe a many-universe system \cite{Strominger90}. Then, it can
represent either: i) a multiverse of parent universes in case that
the nucleated universes are inflating, or ii) a spacetime foam of
continuously creating and annihilating baby universes.

We outline this paper as follows. In sec. II, we derive the
expression for coherent states of a second-quantized universe
using the generalized Heisenberg algebras formalism. In sec. III,
coherent states are computed in quantum cosmology by using a third-quantization description. In section IV, we conclude and add further comments.

\section{Coherent states in the second-quantized multiverse}

In Ref. \cite{GonzalezDiaz07}, a model was considered which provided the second
quantization for a Friedman-Lemaitre-Robertson-Walker (FLRW)
spacetime, filled with an homogeneous and isotropic fluid. The
classical Hamiltonian for that universe is given by,
\begin{equation}\label{classical hamiltonian}
H = - \frac{2 \pi G}{3} a^2 p^2_a + \rho_0 a^{3(1-w)} ,
\end{equation}
where $a$ is the scale factor, $p_a$ is its conjugate momenta, $G$
is the gravitational constant, $\rho_0$ is the energy density of
the fluid at a given time \cite{GonzalezDiaz07}, and $w$ is the
proportionality constant of the equation of state of the fluid, $p
= w \rho$, $p$ and $\rho$ being the pressure and the energy
density of the fluid, respectively. In Eq. (\ref{classical
hamiltonian}), a gauge $N=a^3$ has been used, with $N$ being the lapse function. Then, a set of Hamiltonian eigenfunctions can be
obtained. In the configuration space, they can be written as
\begin{equation}\label{Hamiltonian eigenfunctions}
\phi_n(a) = N_n \mathcal{J}_n(\lambda a^q) ,
\end{equation}
in which $N_n$ is a normalization constant, $\mathcal{J}_n$ is the
Bessel function of the first kind and order $n$, and,
\begin{equation}
q = \frac{3}{2}(1-w) \, \, , \, \, \lambda = \frac{1}{\hbar q}
\sqrt{\frac{3}{2 \pi G} \rho_0}.
\end{equation}
The normalization constants are given by, $N_n = \sqrt{2 q n}$,
for $n>0$. For the zero mode a regularization procedure is needed,
and then \cite{GonzalezDiaz07}, $\frac{N_0^2}{q} \ln
\frac{2}{\lambda l_p^q} = 1$, with $l_p$ some minimal cut-off.
Then, the functions given by Eq. (\ref{Hamiltonian
eigenfunctions}) correspond to the following eigenvalue problem,
\begin{equation}\label{eigenvalue problem}
\hat{H} \phi_n(a) = \mu_n \phi_n(a)  ; \;\; \mu_n = q^2 n^2 ,
\end{equation}
and they are normalized with respect to the scalar product defined
by,
\begin{equation}\label{scalar product general}
<f|g> = \int_0^{\infty} da \, \frac{1}{a} f(a) g(a) ,
\end{equation}
where $\frac{1}{a}$ is a weight factor.

In the case of a dark energy dominated universe, the boundary
conditions that the wavefunctions have to satisfy are
\cite{GonzalezDiaz07}: i) they have to be regular everywhere, even
when the metric degenerates, $a\rightarrow 0$, and ii) they have
to vanish at the big rip singularity when $a\rightarrow \infty$,
in the phantom energy dominated regime. The wavefunctions given by
Eq. (\ref{Hamiltonian eigenfunctions}) obey these boundary
conditions \cite{GonzalezDiaz07}, vanishing as $a \rightarrow 0$, so satisfying the no boundary condition of Hartle and Hawking
\cite{Hartle83}.

Then, a well-defined Hilbert space can be considered, where the Hamiltonian
eigenstates, $|n\rangle$, are those states represented in the
configuration space by the wavefunctions given in Eq.
(\ref{Hamiltonian eigenfunctions}), i.e., $\langle n|a\rangle =
\langle a|n\rangle = \phi_n(a)$, as the wavefunctions considered
so far are real functions. The orthogonality relations for the
Hamiltonian eigenstates can be written then as
\cite{GonzalezDiaz07, RoblesPerez08},
\begin{equation}\label{orthogonality}
\begin{array}{ll}
\langle n | n \rangle = 1 &, \forall n, \\ & \\ \langle n | m
\rangle = 0 & , |n - m| \; \; \;
\textmd{even,} \\ & \\
\langle n | m \rangle = \frac{4}{\pi} \frac{\sqrt{n \, m}}{n^2 -
m^2} & , |n - m|  \; \; \; \textmd{odd}.
\end{array}
\end{equation}
Furthermore, the Hamiltonian eigenfunctions represent valid
semiclassical approximations, i.e., they can be taken to represent
classical universes in the sense that in the semiclassical limit,
$\hbar \rightarrow 0$, they turn out to be quasi-oscillatory
wavefunctions whose argument are essentially given by the
classical action ($S_c = \lambda a^q$). So, the correlations
between the classical variables are satisfied, i.e, $p_a =
\frac{\partial S_c}{\partial a}$ is the classical equation of
motion; and they satisfy also the Hartle criterion
\cite{GonzalezDiaz07, Hartle90}.

Now, we can apply the formalism of generalized Heisenberg algebras
(GHA), such as it is described in Ref. \cite{Hassouni04}, to construct
coherent states for the model being considered. Although coherent states
are usually defined as the eigenstates of the annihilation operator, the GHA procedure allows us to find the coherent states
without knowing the explicit expression of that annihilation
operator. Thus, let us start with a generalized algebra given by,
\begin{eqnarray}
H_0 A^{\dag} & = & A^{\dag} f(H_0) \\ A H_0 & = & f(H_0) A \\
\left[ A^{\dag}, A \right] & = & H_0 - f(H_0) ,
\end{eqnarray}
where $A$, $A^{\dag}$ and $H_0$ are the generators of the algebra,
and $f(x)$ is called the characteristic function of the system.
$H_0$ is the Hamiltonian of the physical system under
consideration, with eigenstates given by
\begin{equation}
H_0 |m \rangle = \mu_m |m\rangle ,
\end{equation}
and $A^{\dag}$ and $A$ are the generalized creation and
annihilation operators,
\begin{eqnarray}
A^{\dag} |m\rangle &=& N_m |m+1\rangle \\ A|m\rangle &=& N_{m-1}
|m-1\rangle ,
\end{eqnarray}
where in our case $N_m^2 = \mu_{m+1} = q^2 (m+1)^2$. The use of a
generalized algebra adds a parametrization through the
characteristic function, $f(H_0)$, that allows us to have a
systematic covering of distinct potentials for the given system.
The customary Heisenberg algebra is recovered in the limiting
value $f(x) = 1+x$ \cite{Hassouni04}.

Then, the coherent states are defined to be the eigenstates of the
generalized annihilation operator,
\begin{equation}
A|z\rangle = z |z\rangle ,
\end{equation}
where $z$ is a generally complex number.

Since we have a Hamiltonian spectrum for the model of a dark
energy dominated universe, (see Eq. (\ref{eigenvalue problem})), we can
now find the characteristic function, $f(x)$, which satisfies
$\mu_{n+1} = f(\mu_n)$ \cite{Curado01}. In the present case, we
have
\begin{equation}
\mu_{n+1} = (\sqrt{\mu_n} + q)^2 \equiv f(\mu_n) .
\end{equation}
The spectrum is formally similar to the spectrum for a free
particle in a square well potential \cite{Hassouni04}, and the
computation to follow can be done in a parallel way. Therefore,
the coherent states are finally given by,
\begin{equation}
|z\rangle = N(z) \sum_{n=0}^{\infty} \frac{z^n}{N_{n-1}!}
|n\rangle ,
\end{equation}
where $N(z)$ is a normalization function of $z$, and
\begin{equation}
N_{n-1}! = q^n n! ,
\end{equation}
with, for consistency, $N_{-1}! \equiv 1$. The coherent states can
then be written as,
\begin{equation}\label{coherent states}
|z\rangle = N(z) \sum_{n=0}^{\infty} \frac{z^n}{q^n n!} |n\rangle
= D(A^{\dag}) |0\rangle ,
\end{equation}
where the displacement operator, $D(A^{\dag})$, is formally given
by
\begin{equation}
D(A^{\dag}) = N(z) I_0 \left( 2 \sqrt{\frac{z A^{\dag}}{q^2}}
\right) ,
\end{equation}
$I_0$ being the modified Bessel function of the first kind of
order zero. In the configuration space, the wavefunctions
corresponding to the coherent states given by Eq. (\ref{coherent
states}) can be expressed in terms of the scale factor, $a$, and
the variable $z$, in the form,
\begin{equation}\label{coherent states configuration}
\langle a | z \rangle = \varphi_z(a) \equiv \varphi(a,z) = N(z)
\sum_{n=0}^{\infty} \frac{|z|^n}{n!} \phi_n(a) ,
\end{equation}
where the function $\varphi(a,z)$ has to be interpreted as a
functional of paths for the scale factor, $a(t)$, and the variable
$z$, which has been re-scaled so that, $\frac{z}{q} \rightarrow z$.

In order to obtain normalized coherent states, it is easier to use
an orthonormal basis for the Hilbert space spanned by the
Hamiltonian eigenfunctions. This can be done by splitting the space
in two parts, corresponding to even and odd modes, respectively,
embedding both in a larger Hilbert space \cite{RoblesPerez08}.
In that case, the normalization functions $N(z)$ can be found,
being
\begin{equation}
|z\rangle = \left( I_0\left( 2 |z| \right) \right)^{-\frac{1}{2}}
\sum_{n=0}^{\infty} \frac{|z|^{n}}{n!} \phi_n(a) ,
\end{equation}
and, then, they satisfy the conditions needed to be a set of
Klauder's coherent states \cite{Hassouni04} (KCS): i)
normalization, ii) continuity in the label $z$, and iii) completeness
\cite{RoblesPerez08}.

On the other hand, these coherent wavefunctions satisfy the
boundary conditions imposed above because they are
satisfied by the Hamiltonian eigenfunctions. When the scale factor
degenerates in the limit $a \rightarrow 0$, by using the asymptotic
expansions for the Bessel functions, we can have for the coherent wavefunctions,
\begin{equation}\label{limit a goes 0}
\varphi(z,a) \approx \frac{1}{\sqrt{I_0(2|z|)}}
\sum_{n=0}^{\infty} \frac{|z|^n}{n!} \frac{\left( \lambda
a^q\right)^n }{2^n n!} = \frac{I_0\left(\sqrt{2 \lambda |z| a^q}
\right)}{\sqrt{I_0(2|z|)}} ,
\end{equation}
which are regular functions, satisfying the
Vilenkin's tunneling condition \cite{Vilenkin86} as it took on a
constant value in this limit.

In the opposite limit, for large values of the scale factor, the
introduced boundary condition is also obeyed. The limit of large
values of the scale factor is equivalent to the semiclassical
limit, where $\hbar \rightarrow 0$. In both cases, the asymptotic
expansions of Bessel's functions are the same, and the Hamiltonian
eigenfunctions go as,
\begin{equation}
\phi_n(a) \approx \sqrt{\frac{2}{\pi \lambda a^q}} \cos\left(
\lambda a^q - \frac{\pi}{2} n - \frac{\pi}{4} \right) .
\end{equation}
Then, the coherent states can be written as,
\begin{widetext}
\begin{equation}\label{coherent wavefunctions}
\varphi (z,a)  \approx  \frac{1}{\sqrt{I_0(2|z|)}}
\sqrt{\frac{2}{\pi \lambda a^q}} \sum_{n=0}^{\infty}
\frac{|z|^n}{n!} \cos\left( \lambda a^q - \frac{\pi}{2} n -
\frac{\pi}{4} \right) = \frac{\cos(|z|- \lambda a^q) - \sin(|z| -
\lambda a^q)}{\sqrt{\pi \lambda a^q I_0(2|z|)}} \rightarrow 0 ,
\end{equation}
\end{widetext}
for large values of the scale factor. Since in this model the
classical action is $S_c = \lambda a^q$, it turns out that the
functional $\varphi(z,a)$ can be also expressed as,
\begin{equation}\label{coherent wavefunctions action}
\varphi (z,a)  \approx  \frac{\cos(|z|- S_c(a) ) - \sin(|z| -
S_c(a))}{\sqrt{\pi S_c(a) I_0(2|z|)}} \rightarrow 0 \; (a
\rightarrow \infty).
\end{equation}
Therefore, we have obtained expressions for normalized coherent
states in the configuration space. They satisfy the imposed
boundary conditions, both, in the limit of large values of the
scale factor and when it degenerates. The same limit for large
values of the scale factor runs for the semiclassical limit, in
which the coherent states should represent, by the Hartle
criterion \cite{GonzalezDiaz07, Hartle90}, valid semiclassical
approximations. That is the case because, for
any value of the parameter $|z|$, Eqs. (\ref{coherent
wavefunctions}) and (\ref{coherent wavefunctions action}) are oscillatory functions of the
classical action with a prefactor which goes to zero as the scale
factor grows up.

\section{Coherent states in the third-quantized multiverse}

Second quantized wavefunctions can describe the quantum state of a
single universe. Furthermore, different Hamiltonian eigenstates
having valid semiclassical approximations can also be considered to describe the state of
parent universes and, in this way, they can be envisaged as a proper representation of the
multiverse. However, the second-quantized theory is physically restricted as it
cannot describe the topological changes associated with the
creation or annihilation of universes. This can be made by using a third-quantization procedure \cite{Strominger90}, in
which a many-universe system can be represented quantum
mechanically. Such a many-universe system can describe either a
multiverse made up of parent universes or a spacetime foam formed by popping baby
universes.

In order to apply the third-quantization procedure to the case of
a set of universes which are dominated by dark energy, let us
start with the second-quantized Hamiltonian given by Eq.
(\ref{Hamiltonian}). We will first show that the states of the
multiverse obtained from different gauge choices of the lapse
function are related to each other by unitary transformations; so,
for simplicity, let us start with the conformal gauge, i.e.,
$N=\frac{3 }{4 \pi G} a $. The Hamiltonian then reads,
\begin{equation}\label{Hamiltonian}
H = - \frac{1}{2} p_a^2 + \frac{1}{2} \lambda_0^2 \, a^{2 (q-1)} .
\end{equation}
The momentum conjugated to the scale factor is now given by, $p_a = -
\dot{a}$, and the action becomes,
\begin{equation}
S = - \int dt  \left(\frac{1}{2} \, \dot{a}^2 + \lambda_0^2 \,
a^{2 (q-1)} \right) .
\end{equation}
The wavefunction of the universe or ground state wavefunction must
satisfy the Hamiltonian constraint, $H \phi_0 = 0$, or if a
canonical quantization is used the Wheeler-DeWitt equation,
\begin{equation}\label{Wheeler-DeWitt equation}
 \frac{1}{2} \ddot{\phi}_0 + \frac{1}{2} \omega^2(a) \phi_0(a) = 0 ,
\end{equation}
where $\omega(a) = \lambda_0 a^{q-1}$  for the case being considered. To third-quantize this second-quantized field theory, we then write an action which is a functional of the second-quantized
wavefunction $\phi(a)$ and reads,
\begin{equation}\label{Action 3-quantized}
^{(3)}S = \frac{1}{2} \int \, da \, \phi\, H \, \phi = \frac{1}{2}
\int \, da \left( \dot{\phi}^2 - \omega^2(a) \phi^2 \right) .
\end{equation}
Variation of Eq. (\ref{Action 3-quantized}) with respect to $\phi$
leads directly to the Wheeler-DeWitt equation (\ref{Wheeler-DeWitt
equation}), and therefore this equation must be assumed to contain all the
information of the second-quantized theory, with the two
formulations being therefore equivalent \cite{Strominger90}. Now, we can
proceed as usual by defining the conjugated momentum, $p_\phi
\equiv \frac{\delta L}{\delta \dot{\phi}}$. The third-quantized
Hamiltonian turns out to be then given by,
\begin{equation}\label{Hamiltonian 3-quantized}
\mathbf{\mathrm{H}} = \frac{1}{2} p_\phi^2 + \frac{\omega^2(a)}{2}
\phi^2 ,
\end{equation}
which is the Hamiltonian for the harmonic oscillator with
\emph{time}-dependent frequency $\omega(a)$. The \emph{time}
variable is now the scale factor, $a$, and therefore the
wavefunction of the multiverse has to satisfy a third-quantized
Schrodinger equation \cite{Strominger90},
\begin{equation}\label{Schrodinger third quantized}
\mathbf{\mathrm{H}} |\Psi \rangle = i \hbar \frac{\partial}{\partial a}
|\Psi \rangle ,
\end{equation}
where $\mathbf{\mathrm{H}}$ is the Hamiltonian of the
third-quantized action, Eq. (\ref{Hamiltonian 3-quantized}). The
meaning of this wavefunction is the following \cite{Strominger90}:
we can decompose $|\Psi\rangle$ at some moment $a$, then
\begin{equation}\label{eq32}
|\Psi \rangle = \sum_N \Psi_N(a) |N> ,
\end{equation}
where $\Psi_N(a)$ is then the probability amplitude for $N$
universes at \emph{time} $a$, or the probability amplitude for $N$
universes with scale factor $a$.

However, Eq. (\ref{Schrodinger third quantized}) is the
Schrödinger equation for an harmonic oscillator with
time-dependent frequency. Harmonic oscillators with time-dependent
mass and frequency have been largely studied in the past
\cite{Lewis69, Pedrosa87}. The wavefunctions can be obtained in
terms of the eigenfunctions of the harmonic oscillator with
constant frequency (i.e., at a given time, $a_0$), because there
is a unitary transformation, $U_\omega$, which in this case turns
out to be a time reparametrization or a reparametrization in the
scale factor, that transforms the harmonic Hamiltonian with time
dependent mass and frequency into the static case \cite{Lewis69}.
Furthermore, the usual creation and annihilation operators for the
harmonic oscillator, $b_0 = \sqrt{\frac{\omega_0}{2 \hbar}} (\phi
+ \frac{i}{\omega_0} p_\phi)$ and $b_0^\dag =
\sqrt{\frac{\omega_0}{2 \hbar}} (\phi - \frac{i}{\omega_0}
p_\phi)$, can be interpreted as the creation and annihilation
operators for the universes, $N_0 = b_0^\dag b_0$ being the number
operator of universes in the multiverse.

In our case, the unitary transformation $U_\omega$ is given by,
\begin{equation}
U_\omega (\phi, a) = e^{-\frac{i}{2 \hbar} \frac{\dot{\rho}}{\rho}
\phi^2} ,
\end{equation}
where,
\begin{equation}
\rho \equiv \rho (a) = \sqrt{\phi_1^2(a) + \phi_2^2(a) } ,
\end{equation}
with,
\begin{equation}
\phi_1(a) = \sqrt{\frac{\pi a}{2 q}} J_{\frac{1}{2q}}\left(
\lambda_0 a^q \right) , \;\; \phi_2(a) = \sqrt{\frac{\pi a}{2 q}}
Y_{\frac{1}{2q}}\left( \lambda_0 a^q \right) ,
\end{equation}
two independent solutions of Eq. (\ref{Wheeler-DeWitt equation}),
with $J_n(x)$ and $Y_n(x)$ the Bessel functions of first and second
kind of order $n$. In that case,
\begin{equation}\label{unitary transformation}
\mathbf{\mathrm{H}} = U_\omega^\dag \, \mathbf{\mathrm{H_0}} \,
U_\omega ,
\end{equation}
where, $\mathbf{\mathrm{H_0}} = \frac{1}{2} (p_\varphi^2 +
\varphi^2)$, is the Hamiltonian for an harmonic oscillator with
constant mass and frequency ($m = \omega_0 = 1$). In obtaining Eq. (\ref{unitary transformation}) the change of
variable $\varphi = \frac{\phi}{\rho}$ has been done. Therefore,
the probability amplitudes for the scale factor-dependent
wavefunctions (\ref{eq32}), are given by
\begin{equation}\label{eigenfunctions}
\Psi_N(a) \equiv \Psi_N (\phi_n, a) = \frac{1}{\sqrt{\rho(a)}} \,
U^\dag_\omega \, \psi_N(\varphi)\mid_{\varphi=\phi} ,
\end{equation}
where $\frac{1}{\sqrt{\rho(a)}}$ is a normalization factor, and
the $\psi_N(\varphi)$ are the eigenfunctions of an harmonic
oscillator with constant mass and frequency, i.e.,
$\mathbf{\mathrm{H_0}} \psi_N = \hbar (N + \frac{1}{2}) \psi_N$.
Thus, any solution of the Schrodinger equation (\ref{Schrodinger
third quantized}) can be written as,
\begin{widetext}
\begin{equation}\label{third-quantized wavefunction}
\Psi(\phi,a) = \sum_N C_N \, e^{i \alpha_N(a)} \left(
\frac{1}{\sqrt{\pi \hbar} 2^N N! \rho} \right)^{\frac{1}{2}}
e^{\frac{i}{2 \hbar}(\frac{\dot{\rho}}{\rho} + \frac{i}{\rho^2})
\phi^2} H_N(\frac{\phi}{\rho \sqrt{\hbar}})
\end{equation}
\end{widetext}
where
\begin{equation}\label{alpha}
\alpha_N(a) = - (N + \frac{1}{2}) \int_0^a \frac{da'}{\rho^2(a')}
.
\end{equation}
The wavefunction given by Eq. (\ref{third-quantized wavefunction})
quantum mechanically represents a general state for a multiverse
made up of flat universes filled with a given homogeneous and
isotropic fluid. The precise kind of such a fluid is encoded in
the potential term of the second-quantized action through the
value taken by the parameter $w$, and hence in the frequency
$\omega(a)$ which appears in the third-quantized action, given by
Eq. (\ref{Action 3-quantized}). The functional form of the
frequency depends thus on the type of fluid which is considered,
i.e., on the type of energy-matter which fills each universe.
However, different solutions for different frequencies of an
harmonic oscillator are related by unitary transformations. In
that sense, the state of the multiverse is invariant to each other
under the kind of matter-energy filling each universe, because
they are states which belong to the same ray in the multiverse.

Therefore, more general potentials could be considered as well as
closed and open geometries for the spacetime. It is thereby more
difficult to compute the solutions of Eq. (\ref{Wheeler-DeWitt
equation}) to obtain the function $\rho(a)$. Nevertheless, the
reasoning used above can be once again applied in a similar way to
the variety of potentials, because the solutions obtained from
different potentials are eventually related by unitary
transformations to those given by Eq. (\ref{third-quantized
wavefunction}). Therefore, the general state for a multiverse made
up of different kind of universes can be written as,
\begin{equation}
|\tilde{\Psi} \rangle = \sum_{\vec{N}} C_{\vec{N}} |N_{1 \omega_1}
\, N_{2 \omega_2} \, \cdots \rangle ,
\end{equation}
where $N_{i \omega_i}$ is the number of universes of type $i$
which correspond to potentials derived from the frequencies,
$\omega_i(a)$.

Furthermore, a conformal time was considered at the beginning of
this section in order to obtain the state of the multiverse. As we
noticed before, other gauge choices could be considered as well.
For a general value of the lapse function, $N \equiv N(a)$, the
second-quantized action in fact reads,
\begin{equation}
S = - \int dt \, N \, \left(\frac{a \dot{a}^2}{2 N^2} +
\lambda_0^2 \, a^{2q-3} \right) ,
\end{equation}
and the corresponding Wheeler-DeWitt equation (\ref{Wheeler-DeWitt
equation}) turns out to be,
\begin{equation}\label{Wheeler-DeWitt equation2}
\frac{N}{a} \ddot{\phi}_0(a) + \lambda_0^2 N a^{2 q - 3} \phi_0(a) = 0 .
\end{equation}
The lapse function, $N$, enters therefore as a mass
term into the equation of motion of the third-quantized harmonic
oscillator because Eq. (\ref{Wheeler-DeWitt equation2}) corresponds to the equation of
a damped harmonic oscillator, i.e.
\begin{equation}\label{damped}
\partial^2_{u u} \phi(u) - \frac{ m'}{m} \partial_u \phi(u) + \omega^2(u) \phi(u) = 0,
\end{equation}
where,
\begin{eqnarray}\label{freq}
\omega^2(u) &=& \left(\lambda_0^2 N a^{2 q -3} \right)_{| a=a(u)} , \\  \label{mas} m(u) &=& \left(\sqrt{\frac{N}{a}}\right)_{| a=a(u)} ,
\end{eqnarray}
$u\equiv u(a)$ being a new variable given by the change, $du =
\sqrt{\frac{a}{N}} da$, and $m' \equiv \partial_u m$, as given in
Eq. (\ref{damped}). The third quantized Hamiltonian corresponds
then to that of a harmonic oscillator with mass and frequency
terms both depending on the scale factor, i.e.
\begin{equation}\label{Hamiltonian 3-quantized2}
\mathrm{H} = \frac{1}{2 m} p_\phi^2 + \frac{m \omega^2}{2} \phi^2 .
\end{equation}
However, under the following canonical transformation,
\begin{eqnarray}
\xi &=& \sqrt{m} \, \phi  , \\ p_\xi &=& \frac{1}{\sqrt{m}} \left(
 p_\phi - \frac{m'}{2} \, \phi \right) ,
\end{eqnarray}
the Hamiltonian given by Eq. (\ref{Hamiltonian 3-quantized2})
transforms into
\begin{equation}
\mathbf{\mathrm{\tilde{H}}} = \frac{1}{2} p_\xi^2 +
\frac{\Omega^2(u)}{2} \xi^2 ,
\end{equation}
and we recover the massless Hamiltonian given by Eq.
(\ref{Hamiltonian 3-quantized}), for a new value of the frequency
given by $\Omega(u) = \Omega(a)_{|a=a(u)}$, with
\begin{equation}
\Omega^2(a) = \omega^2(a) + \frac{1}{4} \partial_a m^2(a) .
\end{equation}
In conformal time, $N=a$ in Eqs. (\ref{freq}) and (\ref{mas}), and
therefore we recover the previous result, i.e. $m = 1$ and
$\Omega(a) = \omega(a) = \lambda_0 a^{q-1}$. However, in terms of
our proper time, for which $N=1$, the new frequency turns out to
be, $\Omega^2(a) = \omega^2(a) - \frac{1}{4 a^2}$, with $\omega^2
= \lambda_0^2  a^{2 q -3}$ and therefore the frequency of the
harmonic oscillator, $\Omega(a)$, diverges when the scale factor
degenerates to zero, a result which is just a consequence from the
chosen gauge.

Thus, quantum harmonic oscillators with time
dependent mass can therefore be eventually related to the solutions of the case of
constant mass and frequency by unitary transformations. Therefore,
the state of the many-universe system is also invariant under the
choice of the lapse function, i.e. under time reparametrizations
inside one of the universes, as it should be expected.

\begin{figure}[h]

\begin{center}

\includegraphics[width=8cm]{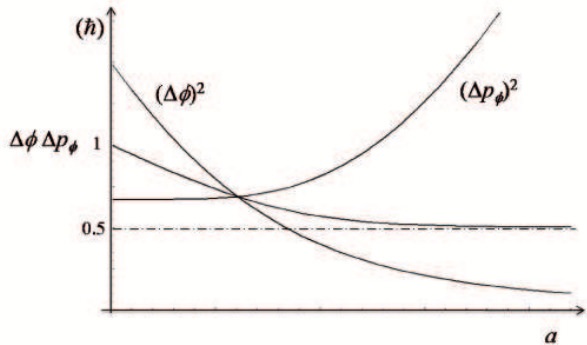}

\end{center}

\caption{$(\Delta\phi)^2$, $(\Delta p_\phi)^2$ and $\Delta \phi
\Delta p_\phi$, for a value $w = - 1$ (vacuum dominated
universes).}

\label{uncertainty1}

\end{figure}

Now, coherent states for the quantum multiverse can be easily
found in the usual way. For the system described by the
Hamiltonian (\ref{Hamiltonian 3-quantized}), the coherent states,
$|\alpha, a \rangle$ read \cite{Pedrosa87}
\begin{equation}
|\alpha, a \rangle = e^{-\frac{|\alpha|^2}{2}} \sum_{N=0}^\infty
\frac{\alpha^N}{\sqrt{N!}} e^{i \alpha_N(a)} |N, a\rangle ,
\end{equation}
where, $|N, a\rangle$ and $\alpha_N(a)$ are given by Eqs.
(\ref{eigenfunctions}) and (\ref{alpha}), respectively. They are
the eigenstates of the annihilation operator, $b(a)$, i.e.,
\begin{equation}
b(a) |\alpha, a\rangle = \alpha(a) |\alpha, a\rangle ,
\end{equation}
where, $\alpha(a) = e^{2 i \alpha_0(a)}$. The scale-factor
dependent annihilation and creation operators are then given by,
\begin{eqnarray}
b(a) &=& \mu(a) b_0 + \nu(a) b_0^\dag , \\ b^\dag(a) &=& \mu^*(a)
b_0^\dag + \nu^*(a) b_0 ,
\end{eqnarray}
where $b_0$ and $b_0^\dag$ are the annihilation and creation
operators of constant mass and frequency ( say, $m = \omega_0 =
1$), and \cite{Pedrosa87}
\begin{eqnarray}
\mu(a) &=& \frac{1}{2} \left( \frac{1}{\rho(a)} + \rho(a) - i
\dot{\rho}(a) \right) , \\ \nu(a) &=& \frac{1}{2} \left(
\frac{1}{\rho(a)} - \rho(a) - i \dot{\rho}(a) \right) ,
\end{eqnarray}
with, $|\mu|^2 - |\nu|^2 = 1$. It follows that coherent states in
the multiverse turn out to actually be describable as squeezed
states \cite{Walls83}. The uncertainty in the wavefunction of a
single universe and its conjugated momentum are in fact given by,
\begin{eqnarray}
(\Delta \phi)^2 &=& \frac{\hbar}{2 \omega_0} |\mu - \nu|^2 ,\\
(\Delta p_\phi)^2 & =& \frac{\hbar \omega_0}{2} |\mu + \nu|^2 .
\end{eqnarray}

\begin{figure}[h]

\begin{center}

\includegraphics[width=8cm]{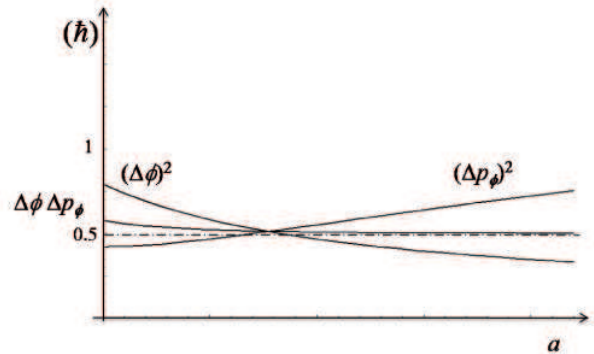}

\end{center}

\caption{$(\Delta\phi)^2$, $(\Delta p_\phi)^2$ and $\Delta \phi
\Delta p_\phi$, for a value $w = 0$ (matter dominated universes).}

\label{uncertainty2}

\end{figure}

\begin{figure}[h]

\begin{center}

\includegraphics[width=8cm]{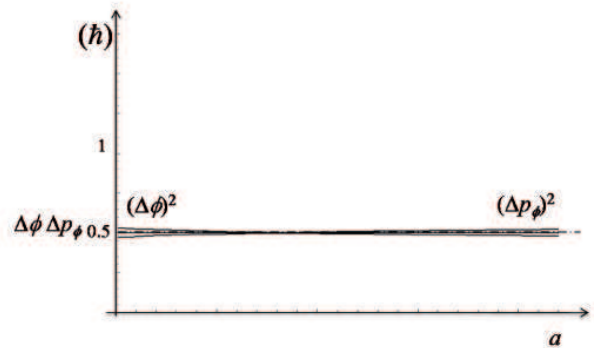}

\end{center}

\caption{$(\Delta\phi)^2$, $(\Delta p_\phi)^2$ and $\Delta \phi
\Delta p_\phi$, for a value $w = 0.3$ (radiation
dominated universes; for a value, $w=\frac{1}{3}$, then,
$(\Delta\phi)^2 =(\Delta p_\phi)^2 = \Delta \phi \Delta p_\phi =
\frac{1}{2}$, and no squeezing effect is present).}

\label{uncertainty3}

\end{figure}

The evolution of such uncertainties are depicted in Figs.
\ref{uncertainty1} - \ref{uncertainty3} for different values of
the parameter $w$. The squeezing effect becomes larger as the
value of $w$ goes away from $\frac{1}{3}$ (i.e., from a radiation
dominated universe), at which point the squeezing effect
disappears, i.e., $(\Delta\phi)^2 =(\Delta p_\phi)^2 = \Delta \phi
\Delta p_\phi = \frac{1}{2}$. Therefore, the squeezing effect
becomes quite more apparent as one is entering in the accelerated
regime of the universe.

\section{Conclusions and further comments}

We have obtained a set of Klauder coherent states for a dark
energy dominated universe. They satisfy the boundary conditions
and can lead to valid semiclassical approximations. Coherent
states represent then a continuous set of states ascribable to
the more classical of probable quantum universes, which are in
this way interpretable as a multiverse. The different universes
residing in such a multiverse differ from one another in a smooth
way by the value taken on by the parameter $z$.

Furthermore, in a quantum multiverse scenario in which topological
changes are allowed to occur, a third-quantization program has
been applied. The state of the multiverse is then obtained in
terms of the eigenstates of an harmonic oscillator with mass and
frequency which depend on the scale factor. The state of the
multiverse is invariant under the energy-matter content of the
universes which form up the whole set, and it is also invariant
under time reparametrizations, as it should be expected.

In the third-quantized description of the multiverse,
coherent states turn out to be converted into squeezed states, the
squeezing effect being larger for accelerated universes. Squeezed
states entail deeper quantum features which have no classical
analogs in the sense that \cite{Reid86} they are described by
non-classical distributions and can violate the Bell's
inequalities, being related therefore with the highly non-local
features of the quantum theory. However, in the context of the
quantum multiverse in which the concept of locality and
non-locality can no longer be applied, these quantum features
would rather be related with the whole universal (independence or
non-independence(?)) mutual interrelation  of the quantum states of
single universes. Therefore, it might well be that accelerated
universes, which are described in the third quantization formalism
by squeezed states, could not be considered as isolated systems
but as really mutually correlated ones within the whole context of
the multiverse, whether or not their quantum states had valid
classical approximations in the semiclassical regime where $\hbar
\rightarrow 0$.

 \acknowledgements This paper was supported by CAICYT
under Research Project No. FIS2005-01181, and by the Research
Cooperation Project CSIC-CNRST.

\end{document}